\documentclass[24pt]{article}
\usepackage{amssymb}
\usepackage{amsfonts}
\usepackage{amssymb,amsmath}
\usepackage{mathrsfs}
\usepackage{latexsym}
\usepackage{amsmath}
\usepackage{latexsym}
\usepackage[cp1251]{inputenc}
\usepackage{graphicx}
\usepackage{color}

\title{Numerical study of quantum mechanical systems using a quantum wave impedance approach}
\author{O. I. Hryhorchak\\
{\small Department for Theoretical Physics, Ivan Franko National
University of Lviv,}\\
{\small 12, Drahomanov Str., Lviv, UA--79005,
Ukraine}\\
\small{\it{Orest.Hryhorchak@lnu.edu.ua}}}

\def\ch{\mathop{\rm ch}\nolimits}
\def\sh{\mathop{\rm sh}\nolimits}
\def\th{\mathop{\rm th}\nolimits}

\begin{document}
\renewcommand{\abstractname}{Abstract}
\maketitle

\begin{abstract}
The approximate numerical method for a calculation of a quantum wave impedance in a case of a potential energy with a complicated spatial structure is considered. It was
proved that the approximation of a real potential by a piesewise constant function is also reasonable in a case of using a quantum impedance approach.The dependence of an accuracy of numerical calculations on a number of cascads by which a real potential is represented was found. The method of including into a consideration of zero-range singular potentials was developed.

\end{abstract}

\section{Introduction}
Numerical calculations are a powerfull tool for a  theoretical investigation of a wide range of quantum mechanical systems. But here we will focus our attention only on multiple-barrier/well systems. These systems were studied numerically by different ways.
For example, in \cite{Babushkin_Nelin:2011} symmetric, asymmet\-ric single- and double-barrier structures as well as the
structure of double-well potential were considered. A particular attention was paid to their characteristics in the region of resonance tunnelling of electrons.
In the paper \cite{Nelin_Imamov:2010} the
features of a numerical Hilbert transform for crystal-like structures are considered and the algorithm
which takes into account these features is offered.
 
For solving multiple-well systems the authors of articles \cite{Calecki_Palmier_Chomette:1984, Tsu_Dohler:1975}.  found the wave functions of each single well and then calculated their overlap within
barriers. But this method is not applicable for thin barriers.
Lui and Fukuma \cite{Lui_Fukuma:1986} derived an analytical solution of a 1D
Shr\"{o}dinger equation for an arbitrary piecewise-linear potential. The authors concluded that since any potential
function can be approximated to arbitrary accuracy using a
piecewise-linear function, the Shr\"{o}dinger equation, in principle, can be solved to any accuracy. But generally this method has similar drawbacks as the use of an approximation by a piecewise constant potential.

One of the most popular techniques in this area is a finite difference method \cite{Zhou:1993, Grossmann_Roos_Stynes:2007}. It allows reducing differential equations to a system of algebraic equations which can be solved by a matrix algebra approach. But very often too many grid points are needed for a necessary accuracy of calculations which make it ineffective for some tasks. Another very effective and widely used technique is a transfer matrix approach
\cite{Ando_Itoh:1987, Griffiths_Steinke:2001, Pereyra_Castillo:2002, SanchezSoto_atall:2012,Harwit_Harris_Kapitulnik:1986,Capasso_Mohammed_Cho:1986,Miller_etall:1985}.

Speaking about numerical calculations on the base of a quantum wave impedance approach one should mention that the paper \cite{Kabir_Khan_Alam:1991} was the first article in which the application of a quantum wave impedance for solving Shr\"{o}dinger equation in quantum wells was considered. It turned out that a technique based on a quantum wave impedance is less restrictive to the shape of a well in a comparison with the other known methods. Developed in that paper method was applied to a parabolic well along with its stepwise approximation. The number of steps was chosen such that the step sizes are much smaller than the electron wavelength. Obtained eigenvalues and eigenfunctions turned out very close to the analytically-computed eigenvalues and eigenfunctions. The issue of using a quantum wave impedance for numerical calculations of chracteristics of quantum systems was the subject of an attention of \cite{Nelin_Imamov:2010, Babushkin_Nelin:2011_1, Ashby:2016}.

The first aim of this paper is to prove that the approximation of a real potential by a piesewise constant function is also reasonable in a case of using a quantum impedance approach and to find the dependence of an accuracy of numerical calculations on a number of cascads by which a real potential is represented. The second aim is to demonstarte (on the example of $\delta$-potential) how to include into a consideration the zero-range singular potentials.

%\section{Approaches for approximate calculations}

\section{Relation between an iterative method and an equation for a quantum wave impedance}
In a paper \cite{Arx3:2020} we have found the iterative procedure of calculating quantum wave impedance for a piecewise constant potential. In this case having a value of a quantum wave impedance function at arbitrary point $x_{arb}$ we can calculate a value of a quantum wave impedance at each point of $x$ axis. But what about an arbitrary non-singular potential? As usual, we can depict it as a sequence of constant potentials with heights $V_i$
and widths $\Delta x_i$. Moreover, in the limit of $\Delta x_i\rightarrow 0$ this depiction coincides with an initial potential.
Thus, on the base of formula for an iterative determination of a quantum wave impedance  \cite{Arx3:2020} we have
\begin{eqnarray}
Z(x-\Delta x_i)=z_i\frac{Z(x)-z_i\th(\gamma_i\Delta x_i)}{z_i-Z(x)\th(\gamma_i\Delta x_i)},
\end{eqnarray}
where $z_i$ is the characteristic impedance of each region of width $\Delta x_i$ and $\gamma_i$ is a wave vector in this region.
If all $\Delta x_i$ are small enough then:
\begin{eqnarray}
Z(x_i-\Delta x_i)=\left(Z(x)-z_i\gamma_i\Delta x_i\right)\left(1+Z(x_i)\gamma_i/z_i\Delta x_i\right)\!+\!
O(\Delta x_i^2)
\end{eqnarray}
or
\begin{eqnarray}
\frac{Z(x_i)-Z(x_i-\Delta x_i)}{\Delta x_i}+Z^2(x_i)\gamma_i/z_i=z_i\gamma_i
+O(\Delta x_i).
\end{eqnarray}
Reminding that $\gamma_i/z_i=\frac{im}{\hbar}$ and $z_i\gamma_i=\frac{2i}{\hbar}(E-U(x_i))$ in the limit of $\Delta x\rightarrow 0$ we obtain the well-known equation for a quantum wave impedance \cite{Arx1:2020}:
\begin{eqnarray}
Z'(x)+\frac{im}{\hbar}Z^2(x)=\frac{2i}{\hbar}(E-U(x)).
\end{eqnarray}
The results of this section give us a base for the approximate numerical calculations of parameters of a studied system with an arbitrary non-singular potential. For this purpose we should depict this potential as a consequence of constant potentials. In the next section we will consider how to use it on practice.

\section{Numerical calculations of a quantum wave \\ impedance using an iterative method}

In a previous section we obtained that in the zero limit of $\Delta x_i$, where $\Delta x_i$ are the widths of regions into which a potential is divided, we get an initial equation for a quantum wave impedance \cite{Arx1:2020}. Now the question is about the most appropriate principle of the division of a potential energy into regions. Usually the potential is divided into regions of an equal width ($\Delta x_i=const$), an equal square ($\Delta x_iU(x_i)=const$) or with an equal parameter $\Delta x_ik_i=const$. The first one is useful in a case of a potential which is localized in a not very wide area. The second and the third one are used when the potential is not a constant in a very wide region. Physically reasonable principle was proposed in \cite{Kabir_Khan_Alam:1991} where the step sizes were chosen to be much smaller than the electron wavelength.

In a bound states case to achieve a desired accuracy $\varepsilon$ one should start from some number $N_0$ of breaking points and to calculate the eigenenergy $E_0$. Then one increases this number to $N_1>N_0$ (very often $N_1$ is chosen to be equal to $2N_0$) and calculate the eigenenergy $E_1$. If $|E_1-E_0|<\varepsilon$ then one stops the process of calculation. If not then one increases the number of breakpoints to $N_2>n_1$ and calculate $E_2$. So, if $|E_2-E_1|<\varepsilon$ one stops the process of a calculation, if not then one continues to increase the number of breakpoints. It is important that theoretically we will definitely achieve the desired accuracy after finite number of cycles because of results of the previous section.

Let's consider the scattering case and assume that we have a wave incidenting on the left of the barrier which has an arbitrary geometry and is described by a function $U(x)$. An approximation of this potential with the piecewise constant potential gives:
\begin{eqnarray}
U(x)&\approx&U_0\theta(x_0-x)+\sum_{i=1}^N U_i\left(\theta(x-x_{i+1})-\theta(x-x_{i})\right)+U_{N+1}
\theta(x-x_{N+1}), 
\end{eqnarray}
where $U_i=(U(x_{i+1})-U(x_i))/2$. Reminding that $\Delta x_i=x_{i+1}-x_i$, $\gamma_i=imz_i/\hbar$  using the relation
\begin{eqnarray}\label{num_iter}
Z_{i-1}={z}_i\frac{Z_j\ch(\gamma_i \Delta x_i)-z_i\sh(\gamma_i \Delta x_i)}
{z_i\ch(\gamma_i \Delta x_i)-Z_i\sh(\gamma_i \Delta x_i)}
\end{eqnarray}
with an initial condition 
\begin{eqnarray}
Z_{N+1}=z_N=\sqrt{2(E-U_{N+1})/m}
\end{eqnarray}
we can calculate $Z_0$. Now having $Z_0$ we are able to find the reflection and transmission coefficients  
\begin{eqnarray}
R(E)=|r(E)|^2,\qquad T(E)=1-|r(E)|^2,
\end{eqnarray}
where a wave reflection amplitude coefficient is equal to 
\begin{eqnarray}
r(E)=\exp[2\gamma_0x_0]\frac{z_0-Z(0)}{z_0+Z(0)}, \quad z_0=\sqrt{\frac{2(E-U_0)}{m}}.
\end{eqnarray}
In a bound states case the process of a calculation is the same but the final relation for the determination of energies of bound states is quite simple:
\begin{eqnarray}
Z(0)=-z_0.
\end{eqnarray}
All that is true only for a non-singular potentials. In a case of a presense of zero-range singular potentials we have to use an another approach, which we will consider in the next section.

\section{Characteristic and input impedances in a case of a $\delta$-potential}

If an approximation of a real potential consists of both a piecewise constant potential and zero-range singular potentials we need to develop a technique which can take into account this fact. In this section we are going to do this using as an example such a zero-range potential as a $\delta$-potential. For this we use the results of papers \cite{Arx1:2020, Arx2:2020, Arx3:2020, Arx4:2020}.

First, let's find an answer for 
a question about the characteristic impedance of the region which includes only one point $x=0$ where $\delta$-well is located. We will  give an answer on this question on the base of a formula:
\begin{eqnarray}\label{Zdelta}
-z_l=z_{\delta_-}\frac{z_l\ch\left(\gamma_{\delta_-}a\right)-z_{\delta_-}\sh\left(\gamma_{\delta_-}a\right)}{z_{\delta_-}\ch\left(\gamma_{\delta_-}a\right)-z_l\sh\left(\gamma_{\delta_-}a\right)},
\end{eqnarray}
where $z_l=Z(a+0)=i\sqrt{2|E|/m}$ is the load impedance, $z_{\delta_-}$ is the characteristic impedance of the region $x\in[0,a], a\rightarrow 0$, $\gamma_{\delta_-}=imz_{\delta_-}/\hbar$. So now we can find the value for $z_{\delta_-}$ from the reduced equation (\ref{Zdelta}):
\begin{eqnarray}
-z_l=z_l-z_{\delta_-}\gamma_{\delta_-}a+o(a),
\end{eqnarray}
or
\begin{eqnarray}
2i\sqrt{2|E|/m}=z_{\delta_-}\gamma_{\delta_-}a+o(a),
\end{eqnarray}
which gives
\begin{eqnarray}\label{z0delta}
z_{\delta_-}=\sqrt{\frac{2\alpha}{ma}}+o(1/\sqrt{a}),\quad a\rightarrow 0.
\end{eqnarray}
The other way to getting the same result for $z_{\delta_-}$ is to consider  a rectangular potential well of a depth $U_0<0$ and of a width $a$. 
Then taking limit 
$a\rightarrow 0,\quad U\rightarrow -\infty$ with assuming $m|U_0|a/\hbar=\alpha>0$ is constant on the base of relation $z_0=\sqrt{2(E+|U_0|)/m}$ we get the same result (\ref{z0delta}).

We can find the characteristic impedance of a region of a $\delta$-barrier in the one of the ways described earlier:
\begin{eqnarray}
\mathop{\mathop{z_{\delta_+}}_{a\rightarrow 0,\: U_0\rightarrow+\infty}}_{maU_0/\hbar=\alpha}\!\!\!\!\!\!\!\!\rightarrow
\sqrt{2(E-U_0)/m}=i\sqrt{\frac{2\alpha}{ma}}+o(1/\sqrt{a}).
\end{eqnarray}

To get an input impedance we use a formula for both a $\delta$-well and a $\delta$-barrier which is similar to (\ref{Zdelta}):
\begin{eqnarray}
z_{i_{\delta_{+,-}}}=z_{\delta_{+,-}}\frac{z_l\ch\left(\gamma_{\delta_{+,-}}a\right)-z_{\delta_{+,-}}\sh\left(\gamma_{\delta_{+,-}}a\right)}{z_{\delta_{+,-}}\ch\left(\gamma_{\delta_{+,-}}a\right)-z_l\sh\left(\gamma_{\delta_{+,-}}a\right)}.
\end{eqnarray}
Substituting obtained expressions for $z_{\delta_+}$  and $z_{\delta_-}$ to this relation we finally get:
\begin{eqnarray}
z_{i_{\delta_-}}=z_l-i\frac{2\alpha}{\hbar},\quad
z_{i_{\delta_+}}=z_l+i\frac{2\alpha}{\hbar}.
\end{eqnarray}
Obtained results correspond to ones in the article \cite{Vodolazka_Nelin:2013}.

How to use these results for numerical calculations in a case of a mix of a piecewise constant potential and $\delta$-potentials? So the answer is as follows. We use the same formula (\ref{num_iter}) as in the previous section everywhere besides points where $\delta$-potentials are located. In the point of a $\delta$-potential location (assume it is point with a number $i-1$, so we can have better connection with a formula (\ref{num_iter})), we replace the value of $Z_{i-1}$ by the value of $Z_{i-1}\pm i\frac{2\alpha}{\hbar}$ that can be depicted as
\begin{eqnarray}
Z_{i-1}\equiv Z_{i-1}\pm i\frac{2\alpha}{\hbar}.
\end{eqnarray}
\begin{figure}[h!]
	\centerline{
		\includegraphics[clip,scale=1.15]{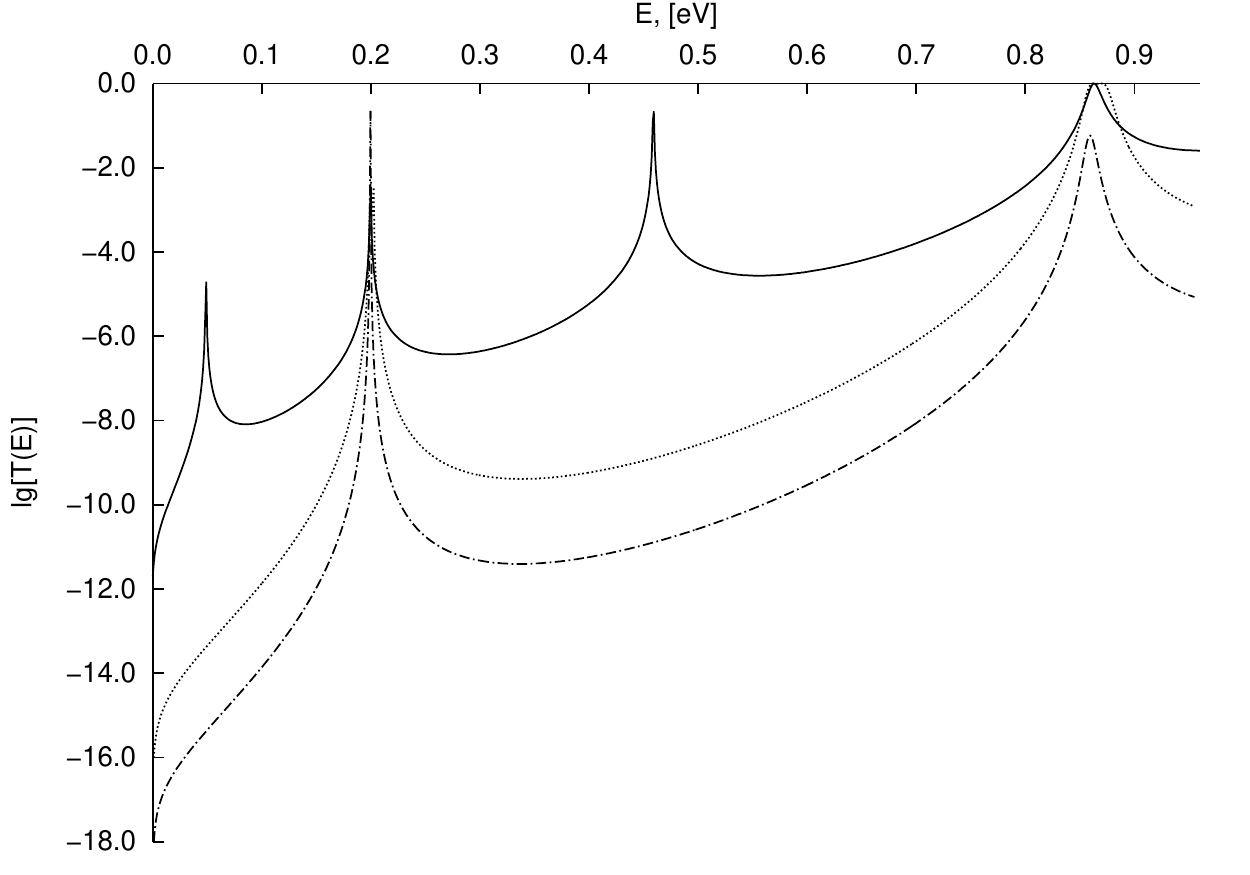}}
	\caption{\small{Dependence of a transmission probability $T$ on an energy $E$ of a particle for a system of rectangular double barrier with a $\delta$-potential $\alpha\delta(x)$ in the middle between the barriers. The height of each barrier is $0.956$ eV. Thickness of each barrier is $30$ nm, distance between them is $100$ nm. Solid line represents the system without a $\delta$-potential ($\alpha=0$); dashed line is for $\alpha=0.25$ keV*nm; dotted line is for $\alpha=0.25$ keV*nm. An effective mass of a particle is $m^*=0.1m_0$, where $m_0$ is a ``bare'' mass of an electron.}}
\end{figure}
Of course it means that if the locations of $\delta$-potentials do not coincide with the locations of interfaces between regions of a piecewise constant potential then the number of breaking points which one use for the approximation of a real potential increases.

Let's illustrate it using the following model:
\begin{eqnarray}
U(x)&=&U_b[\theta(x+a+b)-\theta(x+a)]+\alpha\delta(x)+U_b[\theta(x-a)-\theta(x-a-b)].
\end{eqnarray}
For this case we calculate the transmission probability as a function of an energy. The result is presented on the Figure 1.

\section{Single parabolic barrier. Numerical technique of an iterative calculation}
In this section we are going to illustrate how the numerical technique of an iterative calculation of a quantum wave impedance works in practice. We will consider a scattering case for a single parabolic barrier. A potential energy has the following form:
\begin{eqnarray}\label{sc_par}
U(x)=ax^2\left(\theta(x+x_0)-\theta(x-x_0)\right).
\end{eqnarray}
Parameters $a$ and $x_0$ we define on the base of the paper \cite{Ando_Itoh:1987}. With the help of a technique of an iterative calculation we will numerically calculate the dependence of a transmission coefficient $T$ on an energy $E$ of a particle at different numbers $N$ of breaking points. So, the real potential (\ref{sc_par}) we approximate by the consequence of $N+1$ potential steps. To estimate the accuracy of results which this approximation generates we introduce the average accuracy $\overline{\varepsilon}(N)$ in the following way

\begin{eqnarray}
\overline{\varepsilon}(N)=\frac{1}{n}\sum_{j=1}^n|T^{(N)}(E_j)-T^{(N/2)}(E_j)|,
\end{eqnarray}
where an index $j$ numbers the sequence of energies at which transmission coefficient $T^{(N)}$ is calculated; $n$ is the number of energy values at which $\overline{\varepsilon}(N)$ is calculated. Superscript of a transmission coefficient $T$ shows the number of breaking points at which it is calculated.

On the Figure 2 one can find the dependence of an
average accuracy $\overline{\varepsilon}(N)$ on the number $N$ of breaking points at $n=100$.
\begin{figure}[h!]
	\centerline{
		\includegraphics[clip,scale=1.15]{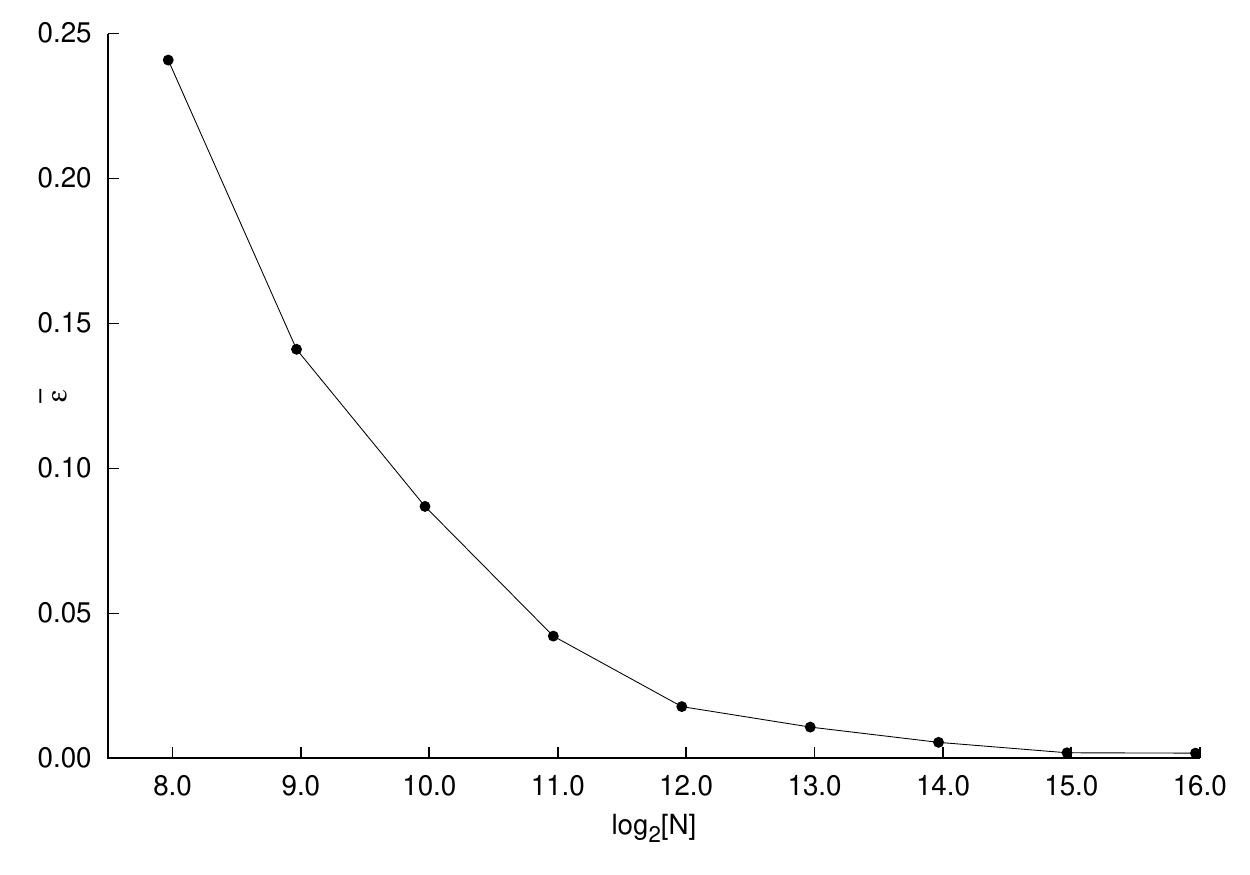}}
	\caption{\small{Dependence of an average accuracy $\overline{\varepsilon}$ on the $log_2(N)$, where $N$ is the number of breaking points. An effective mass of a particle is $m^*=0.1m_0$, where $m_0$ is a ``bare'' mass of an electron.}}
	\label{fig:U1U2U3}
\end{figure}

To consider the bound states case we should slightly modify the previous form of a potential energy:
\begin{eqnarray}
U(x)&=&ax_0^2\left(\theta(-x-x_0)+\theta(x-x_0)\right)+x^2\left(\theta(x+x_0)-\theta(x-x_0)\right).
\end{eqnarray}
The rest steps are the same as in the scattering case.

\section{Conclusions}
The common way of munerical study of quantum mechanical systems with complicated geometry is to represent the real potential by a cascad of constant potentials and then to apply the relevant numerical techniques to the obtained piesewise constant potential. In this paper we  both proved this approach is also relevant in a case of using quantum wave impedance and demonstrated the dependence of an accuracy of numerical calculations on a number of cascads by which a real potential is represented.  

So, if a potential is dived into $N$ regions then a direct using of a Sr\"{o}dinger equation for getting a solution means solving $2N+1$-order determinant, which is not easy task. Applying of transfer matrix technique is helpful but still complicated since it demands multiplying of $N$ $2\times 2$ matrices. 

At the same time the iterative process of a quantum wave impedance calculation demands only $N-1$ cycles.
The effectiveness of this approach is caused by the fact that the equation for a quantum wave impedance  \cite{Arx1:2020} is a first-order differential equation and only one matching condition (instead of two in case of a Sred\"{o}nger equation) is necessary to use at the interface of different regions of a piesewise constant potential. 

All these, along with enriching a technique of a numerical calculation by an introduction of a  zero-range singular potentials, make a significant contribution to the study of quantum mechanical systems and show the usefulness of a quantum wave impedance method for numerical investigation of quantum mechanical systems.
The results of this paper along with \cite{Arx5:2020} can be also applied to the developing a technique of numerical study of infinite and semi-infinite periodic systems.

\renewcommand\baselinestretch{1.0}\selectfont
%\renewcommand{\bibname}{Bibliography} 

%\fancyhead[RE,LO]{\sl Bibliography}

\def\name{\vspace*{-0cm}\LARGE 
	%СПИСОК ВИКОРИСТАНИХ ДЖЕРЕЛ
	Bibliography\thispagestyle{empty}}
\addcontentsline{toc}{chapter}{Bibliography}

{\small

	\bibliographystyle{gost780u}
	%\bibliography{\figsfolder full,add}
	\bibliography{full.bib}
	% insert this in the bbl after	 \begin{thebibliography}{}: \interlinepenalty=10000
	
}

\newpage

\end{document}